# INTRODUCING CHAD - AN ADM1 SOLVER FOR DIRECT LINKING TO LAGRANGIAN CFD SOFTWARE

**Prashant Kumar [1*], Zhenghao Yan[2], Soroush Dabiri[3], Nikolaus Rauch[4] and Wolfgang Rauch[5]**

[1,2,3,5] Unit of Environmental Engineering, University of Innsbruck, Innsbruck, Austria
[4] Interactive Graphics and Simulation Group, University of Innsbruck, Innsbruck, Austria

[1] 0000-0002-5637-1201, prashant.kumar@uibk.ac.at, [2] 0000-0002-5901-3936, zhenghao.yan@uibk.ac.at, [3] 0000-0002-2993-3408, soroush.dabiri@uibk.ac.at, [4] 0000-0003-1326-709X, nikolaus.rauch@uibk.ac.at, [5] 0000-0002-6462-2832, wolfgang.rauch@uibk.ac.at

## Abstract

Standard methods for modeling anaerobic digestion processes assume homogeneous conditions inside the tank and thus suffer from the negligence of hydrodynamics. In this work, we present the software toolbox Coupled Hydrodynamics and Anaerobic Digestion (CHAD), a novel parallelized solver that is capable of utilizing CFD results as the basis for Anaerobic digestion model No.1 (ADMno1) simulations. CHAD uses a particle-based Lagrangian CFD solver i.e., DualSPHysics (DSPH) as input and provides for a parallelized, C++ code implementation of the standard ADMno1. This paper demonstrates a conceptual and numerical verification of the toolbox and outlines the future pathway to enhance the approach.

## 1 INTRODUCTION

Anaerobic digestion (AD) is the eminent method for handling sludge from wastewater treatment and – more general - biochemical degradation of residues from biological waste. AD refers to the biochemical degradation of organic materials in the absence of oxygen resulting in the formation of methane-rich gas. The methane produced can be used as a source of energy and the process results in a reduction of sludge volume, lowering disposal costs. These characteristics make AD an attractive option for sludge treatment as compared to other sludge treatment methods.

Anaerobic digestion model No.1 [1] is the state-of-the-art model for the dynamic simulation of AD processes within a digester. It computes both physio-chemical and biological reactions in the liquid phase and the release of gaseous components into the gas phase. There have been a variety of numerical simulations using ADMno1 to evaluate the performance of an AD tank [2-4]. One major drawback of ADMno1 is that it denotes the digester tank as a continuous stirred tank reactor (CSTR), hence assuming homogeneity of mixed liquor and temperature throughout the tank. This neglects the complexity of spatial distribution of both matter and temperature arising as the consequence of hydrodynamics, thus leading to inaccurate results. CFD, on the other hand, is a well-established numerical tool for simulating hydrodynamics.

It can be safely argued, that the accuracy of AD modeling is improved substantially by incorporating CFD data. However, for doing so one needs to consider the two schools of thought regarding hydrodynamic flow computation i.e., Eulerian and Lagrangian: Eulerian methods discretize the fluid domain into smaller cells (applying a fixed mesh) and solve the resulting equations for each cell. Lagrangian methods, on the other hand, are meshless methods and – for the case of e.g., Smoothed Particle Hydrodynamics (SPH) - divide the volume of fluid into smaller "particles" for simulation. One of the primary benefits of Lagrangian methods, for the simulations of digester tanks, is that the advection of materials is directly handled by the movement of the particles, other than applying the Taylor series to approximate the non-linear term in the transport equation, which brings numerical errors and instability. SPH also allows for direct simulation of free-surface flows. While the benefits of linking particle-based i.e., Lagrangian CFD simulation with biokinetic conversion process models are clear and allow for a refined deterministic representation of the AD process, there are only a few case studies and implementations known that apply integrative models. To our knowledge, only Rezavand et al. [11] and Kumar et al. [12] have applied so far SPH methods in this context. It is more common to apply Eulerian methods in this context as depicted by the examples of Wu et al [5], and Tobo et al.[6], and others. Note that, traditionally CFD is used only to assess the mixing performance while experiments are used to study the effect of the mixers on methane production [7, 8] or solely consider the hydrodynamics [9, 10].

In this work we present the software toolbox Coupled Hydrodynamics and Anaerobic Digestion (CHAD), a novel parallelized solver that is capable of utilizing SPH results as the basis for ADMno1 simulations. DualSPHysics (DSPH) [13], an open-source, C++ and CUDA-based software has been used for providing the hydrodynamic flow conditions. In the following, we briefly present the background of SPH and DSPH as the numerical solver and ADMno1 and its numerical implementation in C++. The core of the paper is about using the flow velocity information for transport and biodegradation of matter as well as the validation of the novel AD model. CHAD is then applied to the simulation of a lab-scale digester as a case study. Finally, we outline the next steps for the further development of CHAD.

## 2 METHODOLOGY

CHAD is a novel software toolbox capable of linking the hydrodynamic data from SPH simulations with the ADMno1 model, hence improving the accuracy of the result by incorporating the effect of fluid behavior. In the following, the background of both SPH (applying the numerical solver DSPH) and ADMno1 is presented first. Next, the concept of integration is depicted i.e., using the SPH particle information for simulating AD particles.

### 2.1 DualSPHysics/SPH

As mentioned earlier, SPH is a meshless method for simulating transient hydrodynamic "problems". Fundamental to the method is the discretization of the fluid domain into (SPH) particles. These (virtual) particles interact with each other and behave based on the solution of discretized momentum and continuity equations.

SPH computes the physical quantity of each particle based on the weighted interpolation of the corresponding quantity of its neighboring particles. For a physical quantity F of particle $i$ the following equation is formulated,

$$F(\boldsymbol{r}_i) = \sum_{j=1}^{N} \frac{m_j}{\rho_j} F(\boldsymbol{r}_j) W(|\boldsymbol{r}_i - \boldsymbol{r}_j|, h) \qquad (1)$$

where $\boldsymbol{r}, m$ and $\rho$ are the position, mass, and density of particles $i$ and $j$. The kernel function, $W$, determines the weighted influence of each particle $j$ within the smoothing radius, $h$, on particle $i$. With this formula, the equations of continuity and momentum conservation can be applied. The other aspects of the SPH methods, such as neighbor search algorithms and artificial viscosity can be found in [13] along with the continuous form of equation (1).

## 2.2 Anaerobic Digestion Model No.1

The AD model and its defining aspects are described below.

**Mass balance, fluid/gas phase**

In the context of ADMno1, an AD tank is assumed to be a continuously stirred tank reactor (CSTR) with an inlet and outlet where the mass balance of each component is computed by the following equation,

$$\frac{dVS_i}{dt} = q_{in}S_{in,i} - q_{out}S_i + V \sum_j \rho_j v_{i,j} \tag{2}$$

where $V$ is the volume of the CSTR, $S_{in,i}$ and $S_i$ are the concentration of component $i$ at the inlet and in the tank respectively, and $q_{in}$ and $q_{out}$ are the flow rates of the inlet and outlet. The kinetic rate $\rho_j$ for process $j$ and biochemical rate coefficients $v_{i,j}$ are based on the stoichiometry table provided by Rosen et al. [14].

**Inhibitions**

ADMno1 takes pH inhibition and other inhibition mechanisms into consideration. The implemented empirical formula for pH inhibition is the Hill equation [14], noted by equations (3) and (4).

$$I_{pH} = \frac{pH^n}{pH^n + K_{pH}^n} \tag{3}$$

$$K_{pH} = \frac{pH_{LL} + pH_{UL}}{2} \tag{4}$$

where $n$ has an empirical value based on Hill functions, $I_{pH}$ is the inhibition coefficient due to pH and $K_{pH}$ is calculated as the average between $pH_{LL}$ and $pH_{UL}$ which are the lower and upper limits of pH inhibition. Some other inhibition mechanisms, namely non-competitive inhibition and substrate limitation, for different reactions are also modeled, by the equations listed below [1],

$$I = \frac{1}{1 + S_I/K_I} \tag{5}$$

$$I = \frac{S_I}{S_I + K_I} \tag{6}$$

where $S_I$ is the inhibitory component and $K_I$ is the inhibition constant. The total inhibition for the processes is a combination of the inhibition mechanism mentioned above.

**Differential Equations and Differential-Algebraic Equations**

Differential Equations (DE) are at the basis of most of the ADMno1 mass balance computations. However, the high stiffness of DE cripples the efficiency when computing fast dynamics like acid-base reaction and mass balance of soluble hydrogen, $S_{h2}$. Instead, for the implementation in CHAD, differential-algebraic equations (DAE) are implemented for those fast reactions to avoid extremely small timestep sizes. The DAE of the acid-base reaction is shown in equation (7).

$$E(S_{H^+}) = S_{cat^+} + S_{nh4^+} + S_{H^+} - S_{hco3^-} - \frac{S_{ac^-}}{64} - \frac{S_{pr^-}}{112} - \frac{S_{bu^-}}{160} - \frac{S_{va^-}}{208} - \frac{K_W}{S_{H^+}} - S_{an^-} \tag{7}$$

With the concentration of the ions of the current time step, the corresponding concentration of protons $H^+$ is solved iteratively based on the equation above, using approaches like the Newton-Raphson method. A similar method is applied to the mass balance computation of $S_{h2}$, as shown by equation (8).

$$E(S_{h2}) = \frac{q_{in}}{V}(S_{h2,in} - S_{h2}) + \rho_{S_{h2}} - \rho_{G_{h2}} \tag{8}$$

where $\rho_{S_{h2}}$ is the rate of uptake of soluble hydrogen and $\rho_{G_{h2}}$ is the production rate of the gas phase hydrogen.

### 2.3 CHAD Concept

CHAD is a novel software toolbox that provides a dynamic ADMno1 simulation based on spatial flow dynamics. As depicted in Figure 1, CHAD represents each fluid (SPH) particle by an equivalent AD particle in the form of a CSTR tank. CHAD simulates the same number of AD tanks (denoted as AD particles) as SPH particles. Consequently, the volume of the AD particles is determined by the total tank volume divided by the number of particles.

The idea is that – for a given timestep - the velocity, position, and density information of all the particles of the SPH simulation are loaded to the CHAD toolbox. Thus, the spatial displacement of each AD particle for the timestep is determined. Other than for Eulerian approaches advection does not need to be considered for the transport of substances as being taken care of by the Lagrangian approach implicitly (by the movement of the particles). What must be considered still is diffusion. However, note that at this point we refrained from implementing diffusion but test CHAD for advective transport only (see our remarks in the section on future developments on this issue).

Once the spatial information is received the bio-kinetic simulations are carried out for that time step for each AD particle. Note that the internal time steps of the CFD computation do not need to be similar to the ADMno1 simulation but only synchronized. As compared to SPH hydrodynamics the timesteps needed for stable ADMno1 computation are orders of magnitudes larger, so time step management is necessary. At this point, we are applying synchronized timesteps for the CHAD computation of appr. 1 second to limit spatial displacement of the AD particles in this period. For such small timesteps, we additionally do not need to concern ourselves with adaptive timestep management for solving the differential equations arising from the ADMno1.

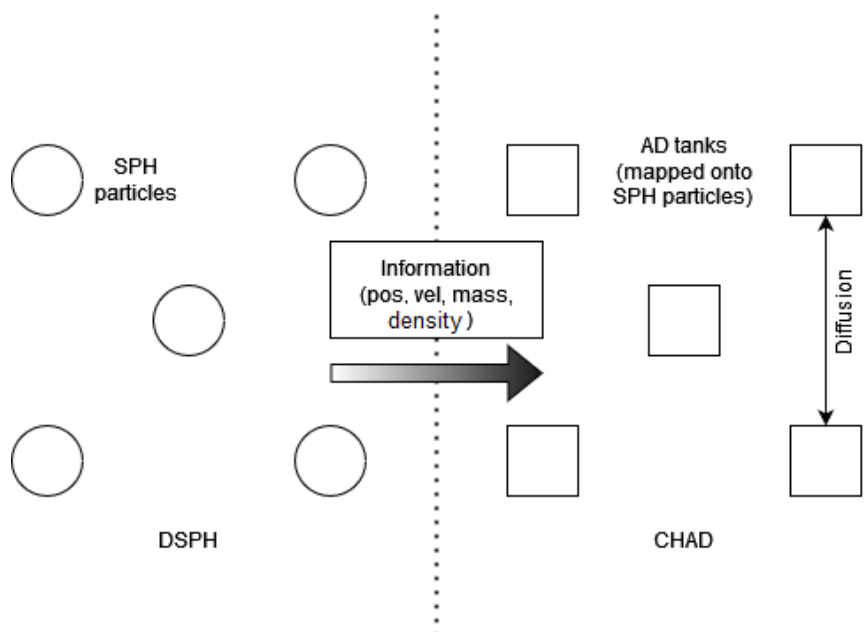

*Figure 1: Overlaying ADMno1 onto SPH data using CHAD*

## 3   CASE SETUP

At this point, we aim to validate and test the current numerical implementation of CHAD, rather than to demonstrate a wide range of possible applications. Thus, the two case studies are selected for validation of the concept and numerical verification of the solver.

### 3.1   Case 1 – Single-particle simulation

CHAD is run as a conventional ADMno1 solver with a single particle that emulates a CSTR. This case is used solely to test the ADMno1 solver and to verify the results with the benchmark implementation by Rosen et al. [14]. The properties of the tank/ particle are given in Table 1. The particle/CSTR is initialized with the concentrations as provided within the BSM2 framework. However, note that the benchmark model [14] does not include a separate ADMno1 but instead a whole wastewater treatment plant model i.e., a combination of ASM and ADMno1 simulations. Consequently, the boundary conditions to their ADMno1 implementation – most important the inflow – contain multiple steps which are not incorporated in CHAD. This is because, currently, CHAD is intended to be used in conjugation with SPH only to study the combined effect of hydrodynamics and biochemical reactions. Thus, CHAD cannot reproduce exactly the results from the BSM2 framework but close approximations only.

*Table 1. Paraments for CSTR simulation (Case 1)*

| Parameter | Value |
|---|---|
| Fluid Volume | 3400 $m^3$ |
| Gas Volume | 300 $m^3$ |
| Inflow/Outflow | 178.45 $m^3$/day |
| Temperature | 308.5 K |
| Atmospheric pressure | 1.013 bar |

*Table 2. Paraments for lab-scale digester (Case 2)*

| Parameter | Value |
|---|---|
| Volume of tank | $8 \times 10^{-3}$ $m^3$ |
| Mixer Velocity | 12 rpm |
| Simulation Duration | 200 sec |
| Temperature | 308.5 K |
| Number of Particles | 128726 |
| Volume of each particle | $6.4e^{-8}$ $m^3$ |

### 3.2   CASE 2- Lab-scale Digester simulation

In the second case study we use a lab-scale digester as shown in Figure 2. This reactor is used for lab-scale experiments representing a real-world digester (see Neuner et al. [15]) and the features are presented in Table 2. This tank has neither an intake nor an outflow. Fluid is added to the tank before experiments and a helical mixer is used to provide mixing. This digester is made out of acrylic material to allow for the use of PIV methods to study the hydrodynamics. Due to this, sludge cannot be added to the digester as PIV techniques would fail. Although useful for hydrodynamics, it gives no information about the biokinetics expected.

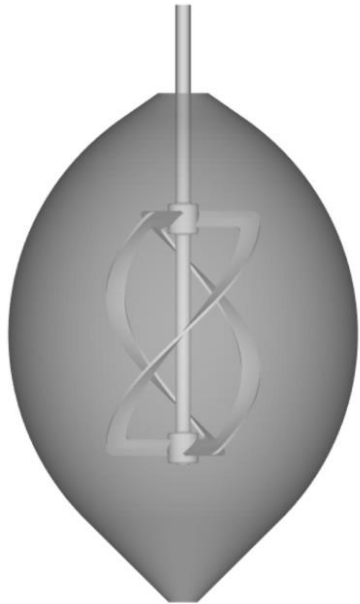

*Figure 2: Lab-scale model of AIZ digester with helical mixer*

At this point, there is not sufficient information available to allow for a direct comparison of the CHAD simulation with the measured concentrations of substances and biogas release. Thus, we use this case study predominantly to test for the general concept to apply the spatial information of N number of AD particles and perform the biokinetic calculations. We further use the case study to test computational time issues and parallelization.

## 4    RESULTS AND DISCUSSION

### 4.1    CASE1- single particle, CSTR

As mentioned, the presented simulation is used as a numerical validation for the ADMno1 implementation within CHAD. For this purpose, the results obtained are compared with the benchmark models obtained by Rosen et. al. [14]. Their model is a C++ integrated with MATLAB code which considers most of the processes within a WWTP. We, however, compare only the results of their AD modeling.

Fig. 3 shows the concentration of various components which are either produced or utilized at each of the stages of the AD process within 60 days. The organic biomass is broken down during disintegration resulting in the formation of carbohydrates, the concentration of which is shown in Fig 3a. This is then hydrolyzed to form sugars (Fig 3b). The concentration of butyrate is shown in Fig 3c which is of the three volatile fatty acids (VFA) produced as a result of acidogenesis. The VFA undergoes acetogenesis to produce acetate (Fig 3d) which finally undergoes methanogenesis to form methane. The concentration of both the soluble and gaseous forms of methane is shown in Fig 3e and Fig 3f respectively.

The concentrations obtained via CHAD are also compared to the benchmark model in Fig 3. It can be inferred that the results from both models are within an acceptable range of each other. The slight discrepancies can be attributed to the varying inflow conditions for the CSTR, due to the inclusion of a variety of other processes (such as the Active sludge model, dewater and clarifier). However, the inflow concentration in CHAD is assumed to be constant.

Kumar, Yan & Rauch (2022)

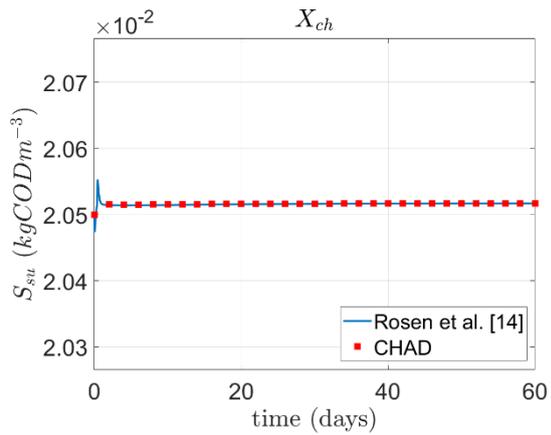
*Figure 3 a Concentration of carbohydrates, $X_{ch}$*

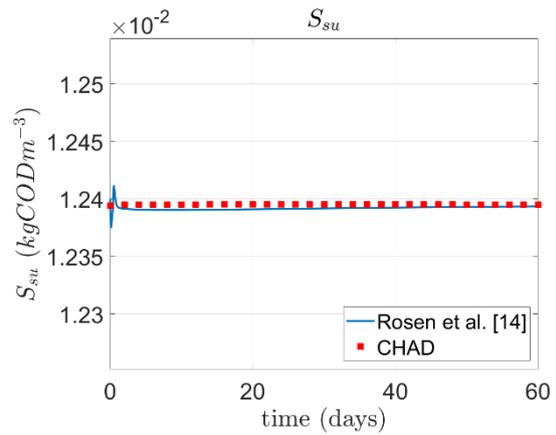
*Figure 3 b Concentration of soluble sugars, $S_{su}$*

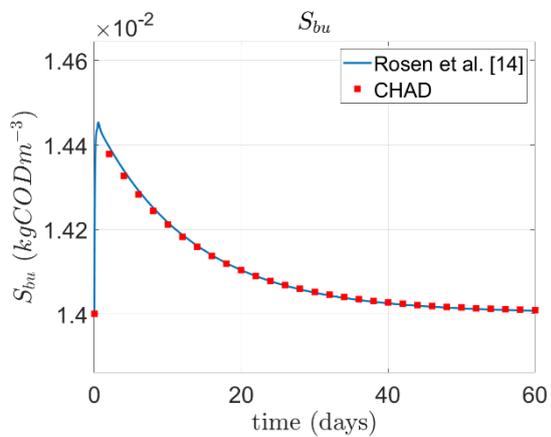
*Figure 3 c Concentration of butyrate, $S_{bu}$*

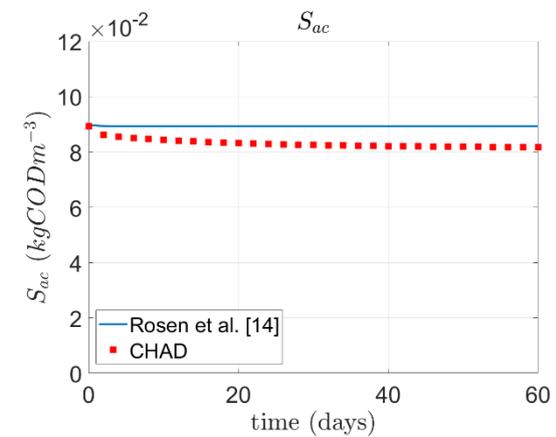
*Figure 3 d Concentration of acetate, $S_{ac}$*

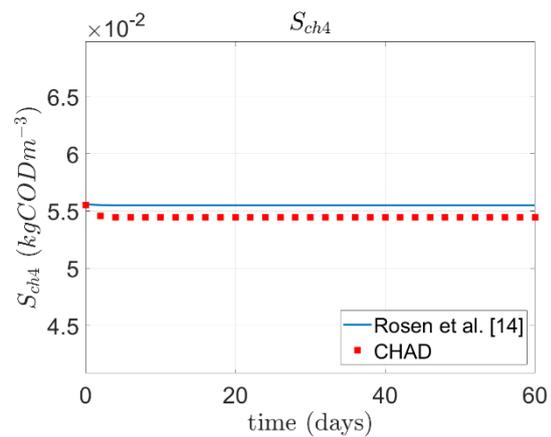
*Figure 3 e Concentration of soluble methane, $S_{ch4}$*

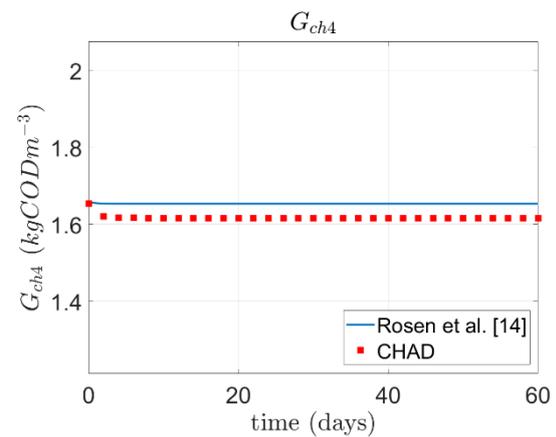
*Figure 3 f Concentration of gaseous methane, $G_{ch4}$*

*Figure 3: Comparison of ADMno1 component concentration between benchmark model and CHAD over 60 days, where $X_{ch}$ is the insoluble carbohydrates, $S_{su}$ is the soluble monosaccharides, $S_{bu}$ is the soluble butyrate, $S_{ac}$ is the soluble acetate, $S_{ch4}$ is the soluble methane and $G_{ch4}$ is the methane in the gas phase.*

The pH of the sludge plays an important role in regulating the reaction rate of processes. As shown in Fig.4, the pH value is fairly consistent with the benchmark model, considering its high sensitivity.

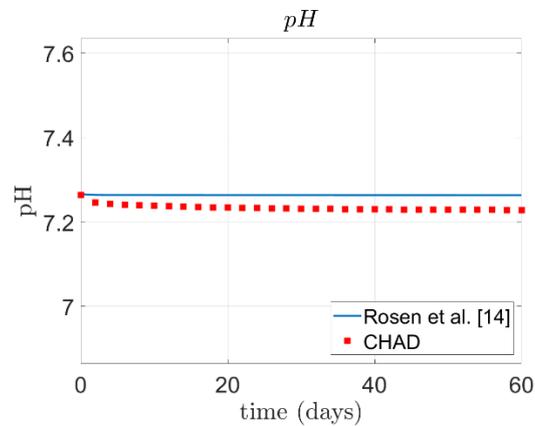

*Figure 4: Comparison of pH between benchmark model and CHAD over 60 days*

The error analysis of biomass ($X_c$), each of the components shown in Fig.3 and pH are listed in Table 3, shown below.

*Table 3. Relative root mean square error of the ADMno1 components and pH*

| Parameter | Error (%) |
|---|---|
| Biomass, $X_c$ | 0.09 |
| Carbohydrates, $X_{ch}$ | 2.07e$^{-3}$ |
| Sugars, $S_{su}$ | 0.02 |
| Butyrate, $S_{bu}$ | 0.03 |
| Acetate, $S_{ac}$ | 7.80 |
| Soluble methane, $S_{ch4}$ | 1.88 |
| Gaseous methane, $G_{ch4}$ | 2.22 |
| pH | 0.45 |

## 4.2 CASE2 – Lab-scale simulations

This simulation shows the ability of CHAD to be coupled with the results of an SPH software. As mentioned in the previous sections, the spatial data of the particles was provided to CHAD, which was able to effectively overlay ADMno1 over it. CHAD then calculated the changes in concentrations of the substrate within the tank over time. The iso-surface representing velocities that are obtained from the SPH simulation and the concentration of methane can be seen in Fig 5 and Fig 6 respectively. As diffusion is yet to be implemented in CHAD and the initial conditions of all the particles are the same, the chemical compositions of all the particles are similar.

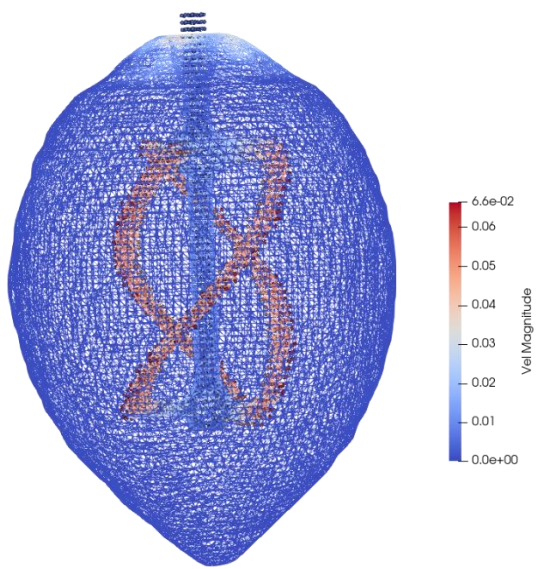

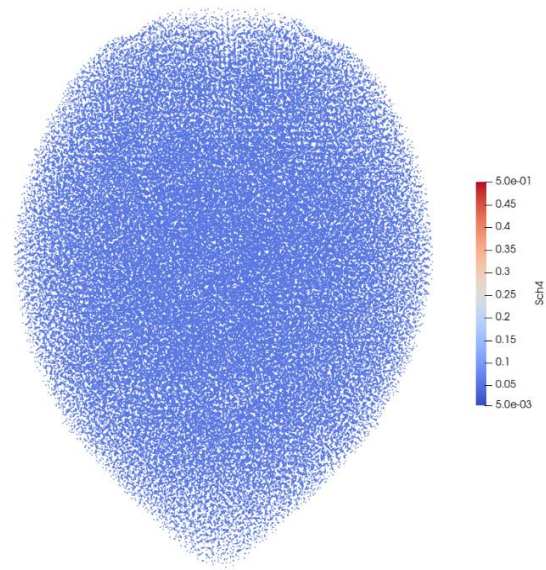

*Figure 5: Velocity magnitude (m/s) from DSPH*

*Figure 6: Dissolved methane concentration (kgCOD/m³)*

### 4.3 Parallelization

As CHAD computes the ADMno1 components independently for each particle it lends itself to parallelization i.e., processing particles simultaneously on different processors. To this end, the workload is divided into equal chunks (number of particles) and distributed to multiple threads.

We tested the implementation on a Desktop PC with an AMD Ryzen 7 3700X 8-Core@3.6GHz Processor and 48 GB DD3 RAM. Specifically, the runtime of processing an increasing number of particles with a different number of threads is compared for one timestep Dt = 0.5 sec with 10 internal steps for the ADMno1 computation $dt$=0.05 sec. Figure 1 depicts the mean runtime of ten simulations to account for numerical effects. As expected, due to the independent computation, the current implementation of CHAD has a linear runtime complexity with respect to the number of particles. In the end, a speed-up of $\approx 7.82$ (or an 87% reduction in computation time) for simulating the ADMno1 reactions for fluid particles of the CSTR simulation (110929 Particles) is achieved.

Note, that this is the computation time of the ADMno1 concentration for all particles – without reading the SPH particle data from a file and writing the results ($\approx 4s$ and $\approx 7.5s$ respectively for 1M particles, both ASCII files). Therefore, file IO is one of the main bottlenecks of the current implementation, which will be improved in the future by switching to binary files.

With the introduction of diffusion between particles (see Future Developments), which will be based on an SPH-based discretization of the diffusion equation, a sophisticated nearest neighbor search is required – due to the brute force approach having quadratic runtime complexity. To this end, we will implement a parallelized hash-grid nearest neighbor search.

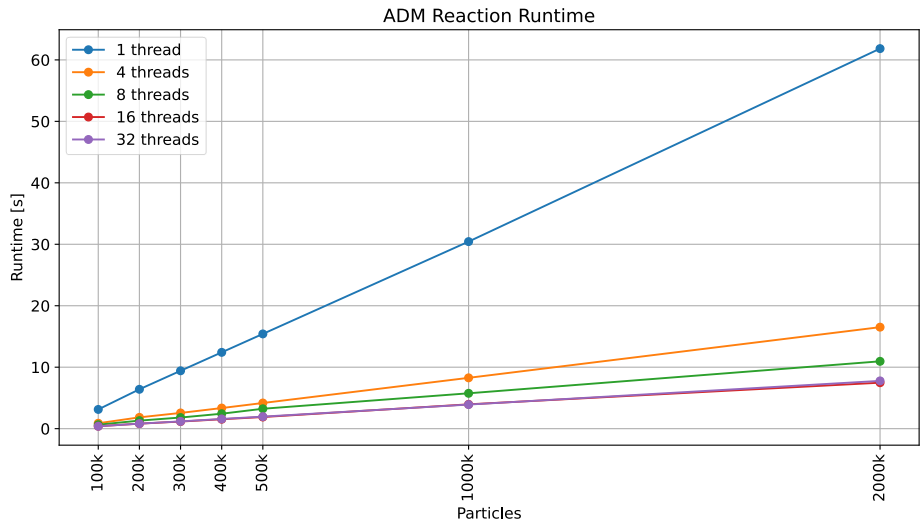

Figure 7: Runtime of CHAD-ADMno1 for one timestep

## 5 CONCLUSIONS AND FUTURE DEVELOPMENT

CHAD is a software toolbox to provide for a coupled hydrodynamic – biokinetic simulation of AD tanks for (particle-based) Lagrangian CFD solvers. The stand-alone AD model within CHAD has been verified with the industry-standard benchmark model and it shows minor variations. CHAD used is highly parallelizable due to its independent particle approach resulting in fast computations with capable hardware.

At this point CHAD is limited to advection transport processes but will be enhanced as follows:

- Diffusion: The key step and necessary for practical applications in the implementation of the diffusion process in the toolbox. This does not pose any theoretical problems but is primarily a numerical problem. For diffusion in a given timestep all AD particles in the vicinity of each other need to be considered. For practical purposes, we aim to implement a neighborhood search with the SPH kernel framework (see e.g., Winkler et al., [16]).

- Open CFD input: SPH has been developed for rapid changes in hydrodynamics, and this is not well suited for slow-moving fluids as given in AD tanks. The virtue of SPH is that it is both a well-established and a thoroughly validated method with a rigorous mathematical concept. Still, the concept of CHAD is likewise applicable to other – more simplified and thus much faster – particle-based methods such as FLIP and alike. We aim to implement the appropriate interfaces.

- Multiple phases: For practical purposes the consideration of sediments and non-biodegradable substances in the reactor is important. This both influences mixing processes as well as shapes the flow field eventually filling the tank starting from the bottom. The CHAD concept allows to consider multiple phases both in the hydrodynamic computation but also in the biokinetic conversion processes by introducing different types of particles.

- Biokinetic models: at this point, CHAD includes ADM1 as a biokinetic model. However, the concept can be applied to any problem where the integration of hydrodynamics with a biokinetic conversion process is meaningful. Typical examples are found in SBR reactors or denitrification tanks in AOB wastewater treatment, where mixing is an important element of the process.

## ACKNOWLEDGMENT

This research is part of a project which is supported by the Federal Ministry of the Republic of Austria for Agriculture, Regions and Tourism in collaboration with the Kommunalkredit Public Consulting GmbH [Grant Number: B801259].

Zhenghao Yan has received funding from the European Union's Horizon 2020 research and innovation program under the Marie Skłodowska-Curie grant agreement No. 847476. The contents of this publication do not necessarily reflect the position or opinion of the European Commission